\documentclass[reprint,aps,prb,showpacs]{revtex4-1}

\usepackage{graphicx}
\usepackage{dcolumn}
\usepackage{bm}
\usepackage{hyperref}
\usepackage{siunitx}

\graphicspath{{./figures/}} 

\begin{document}
\title{Spin relaxation and coherence times for electrons at the Si/SiO$_2$ interface}

\author{S. Shankar}
\altaffiliation[Current address : ]{Applied Physics Dept., Yale University, New Haven, CT 06511, USA}
\email{shyam.shankar@yale.edu}

\author{A. M. Tyryshkin}

\author{Jianhua He}

\author{S. A. Lyon}

\affiliation{
Dept. of Electrical Engineering, Princeton University, Princeton, NJ 08544, USA
}

\begin{abstract}
While electron spins in silicon heterostructures make attractive qubits, little is known about the coherence of electrons at the Si/SiO$_2$ interface. We report spin relaxation (T$_1$) and coherence (T$_2$) times for mobile electrons and natural quantum dots at a $^{28}$Si/SiO$_2$ interface. Mobile electrons have short T$_1$ and T$_2$ of \SI{0.3}{\micro\second} at \SI{5}{\kelvin}. In line with predictions, confining electrons and cooling increases T$_1$ to \SI{0.8}{\milli\second} at \SI{350}{\milli\kelvin}. In contrast, T$_2$ for quantum dots is around \SI{10}{\micro\second} at \SI{350}{\milli\kelvin}, increasing to \SI{30}{\micro\second} when the dot density is reduced by a factor of two. The quantum dot T$_2$ is shorter than T$_1$, indicating that T$_2$ is not controlled by T$_1$ at \SI{350}{\milli\kelvin} but is instead limited by an extrinsic mechanism. The evidence suggests that this extrinsic mechanism is an exchange interaction between electrons in neighboring dots. 
\end{abstract}

\maketitle

\section{Introduction}
\label{sec:introduction}

Electron spins confined in quantum dots in silicon are a promising quantum computing architecture\cite{Loss1998,*Vrijen2000,*Friesen2003,Hanson2007} that builds on the substantial experience with scaling classical electronics. One challenge for these solid-state based qubits is that the spin relaxation (T$_1$) and coherence (T$_2$) times need to be much longer than the gate operation time, to ensure that errors do not destroy a computation\cite{DiVincenzo2000}. Silicon based heterostructures are particularly advantageous in this regard since spins in silicon can potentially have long T$_1$'s and T$_2$'s. For example, extensive measurements have shown that electron spins bound to donors in bulk silicon have long relaxation times\cite{FeherGere1959,*Gordon1958,*Chiba1972}, with coherence times reaching over \SI{1}{\second} at \SI{1.8}{\kelvin}\cite{[{Recent unpublished results. A. M. Tyryshkin, S. A. Lyon, S. Tojo, K. M. Itoh, J. J. L. Morton, M. L. W. Thewalt, H. Riemann, N. V. Abrosimov, P. Becker, H.-J. Pohl, (unpublished). The longest published T$_2$ is \SI{60}{\milli\second} at \SI{7}{\kelvin}. See }]Tyryshkin2003}. In contrast, only a few measurements of T$_1$ have been reported on spins confined in quantum dots in silicon~\cite{Hayes2009,Xiao2010} while there have been no measurements of T$_2$. Thus, for progress towards a practical technology, it is crucial to measure the T$_1$ and T$_2$ of spins confined in silicon quantum dots as well as to understand the mechanisms limiting them.

Here we perform electron spin resonance (ESR) experiments to measure the T$_1$ and T$_2$ of mobile 2D electrons as well as electrons confined in natural quantum dots in a silicon MOSFET. We find that mobile electrons at the Si/SiO$_2$ interface have a T$_1$ and T$_2$ of about \SI{0.3}{\micro\second} at \SI{5}{\kelvin}, shorter than for 2D electrons in Si/SiGe heterostructures~\cite{Tyryshkin2005}. As in Si/SiGe structures, spin relaxation may be controlled by a fluctuating Rashba effective magnetic field (B$_{BR}$) due to the spin-orbit interaction~\cite{BychkovRashba1984,Tahan2005}. 

We have previously~\cite{Shankar2008} shown that when the MOSFET gate voltage is lowered below threshold, the natural disorder at the Si/SiO$_2$ interface fortuitously confines electrons into isolated, independent dots with confinement energies of a few millivolts. While not obviously useful for quantum information processing, these natural dots provide a system to study spin decoherence processes that might affect gate defined quantum dots. Here we extend our previous experiments to show that T$_1$ for some of these natural dots increases rapidly with decreasing temperature below \SI{1}{\kelvin}, reaching \SI{0.8}{\milli\second} at \SI{350}{\milli\kelvin}. Our results are consistent with theory~\cite{Khaetskii2001} and other recent experiments in silicon~\cite{Hayes2009,Xiao2010} showing that upon confinement and cooling to low temperature, the Rashba field is less effective in inducing the relaxation. 

The longest T$_2$ measured for the natural dots is \SI{30}{\micro\second} at \SI{350}{\milli\kelvin}, two orders of magnitude longer than for mobile electrons but shorter than the T$_1$ of \SI{0.8}{\milli\second}. T$_2$ depends on the density of confined electrons, with shorter T$_2$ observed for higher dot densities. This observation suggests that exchange interactions between the dots is the limiting mechanism for T$_2$ which may explain why $\mathrm{T}_2 \ll \mathrm{T}_1$.

\section{Materials and Methods}
\label{sec:Materials}

A silicon n-channel accumulation MOSFET was fabricated on an isotopically enriched $^{28}$Si epi-wafer having \SI{800}{ppm} of $^{29}$Si. The (100) oriented epilayer was \SI{25}{\micro\meter} thick and was doped with phosphorus to a density of \SI{e14}{\per\cmc}. The MOSFET had a large gate area ($\num{0.4}\times\SI{2}{\cms}$) in order to obtain adequate ESR signal from 2D electrons~\cite{Shankar2008}. The device consisted of phosphorus implanted source-drain contacts, a \SI{110}{\nm} dry thermal gate oxide, and a Ti/Au metal gate. Transport measurements at \SI{4.2}{\kelvin} gave a threshold voltage of \SI{1}{\volt}, an electron density of \SI{2e11}{\per\cms\per\volt} and a peak Hall mobility of \SI{14000}{\cms\per\volt\per\second}.

Continuous wave (CW) and pulsed ESR experiments were performed using a commercial ESR spectrometer (Bruker Elexsys580) operating at X-band frequencies (\SIrange[tophrase=dash,repeatunits=false,trapambigrange=false]{9}{10}{\GHz}). A $^{3}$He cryostat (Janis Research) was used to maintain temperature down to \SI{350}{\milli\kelvin}. The spin dephasing time (T$_2^*$) caused by static inhomogeneities was extracted from the linewidth measured in the CW ESR experiment. Standard pulse sequences were used to measure T$_1$ and T$_2$\cite{Schweiger2001}. T$_2$ was measured using a 2-pulse Hahn echo sequence ($\pi$/2 -- $\tau$ -- $\pi$ -- $\tau$ -- echo). T$_1$ was measured using a 3-pulse inversion recovery sequence comprising an initial inverting $\pi$ pulse followed by a measurement sequence using 2-pulse echo detection ($\pi$ -- t -- $\pi$/2 -- $\tau$ -- $\pi$ -- $\tau$ -- echo). Imperfections in the microwave pulses can contaminate Hahn echo and inversion recovery decays with extraneous signals that decay on a timescale of T$_2^*$. Therefore, a 16-step phase cycling sequence was used to remove these extraneous signals and measure the true T$_2$ and T$_1$ (Ref.~\onlinecite{Schweiger2001}, p. 133-135). 

Microwave pulses of optimal power (\SI{520}{\ns} $\pi$ pulse length), were used to avoid electron heating. At higher powers, T$_1$ for natural dots decreases indicating some microwave heating. Unfortunately, below this power, the microwave magnetic field B$_1$ was smaller than the ESR linewidth (\SI{20}{\micro\tesla}) resulting in reduced sensitivity. While we cannot rule out microwave heating at the lowest temperatures, the pronounced temperature dependence of T$_1$ suggests that residual heating is small. 

\section{Results}
\label{sec:Results}

\begin{figure}
\centerline{\includegraphics*{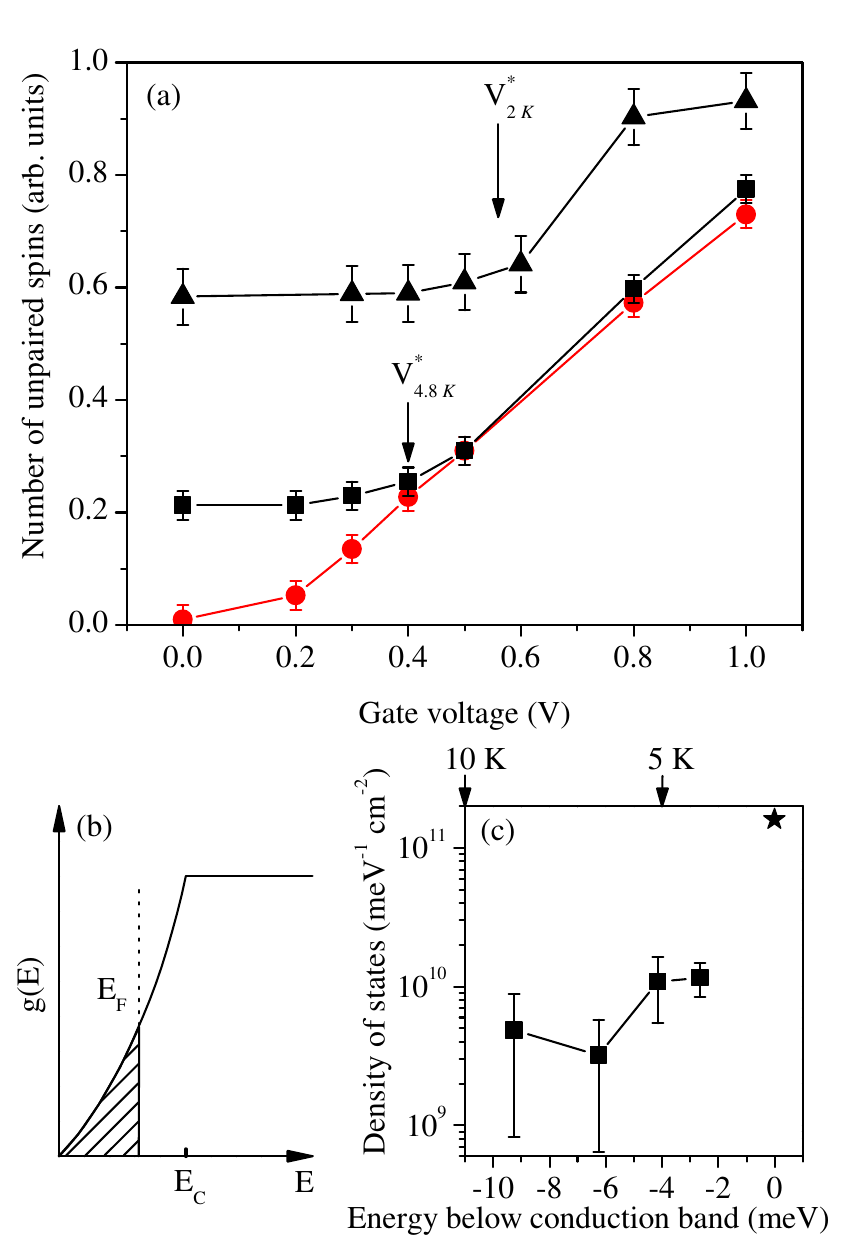}}
\caption{\label{fig:ESR signal vs gate voltage}(a) Number of unpaired spins (N$_S$) as a function of gate voltage (V$_G$) below threshold ($\mathrm{V}_{th} =\SI{1}{\volt}$) as measured at \SI{4.8}{\kelvin} in the dark (squares) and after illumination (circles), and at \SI{2}{\kelvin} in the dark (triangles). Arrows on the curves recorded in the dark indicate a characteristic voltage V$^*_T$ at which N$_S$ becomes constant while decreasing V$_G$ from V$_{th}$. Lines are guides for the eye. (b) Cartoon of density of states, g(E), of 2D electrons; g(E) is constant above the conduction band edge (E$_C$) and decays below E$_C$. States below E$_F$ are occupied. In the experiment, E$_F$ can be adjusted by the gate voltage bias V$_G$. (c) Density of states of confined electrons as a function of energy below E$_C$, as obtained from data in Fig.~\ref{fig:ESR signal vs gate voltage}(a). The star indicates the calculated value (\SI{1.6e11}{\per\meV\per\cms}) of the density of states in the conduction band for 2D electrons at a Si(100) surface (two-fold valley degeneracy). The labeled arrows on the top axis indicate the depth of confined electrons after setting $\mathrm{V}_{G} = \SI{0}{\volt}$ in the dark at \SI{5}{\kelvin} and \SI{10}{\kelvin}, respectively.}
\end{figure}

The ESR spectra of the MOSFET reveal a gate voltage (V$_{G}$) dependent signal having a g-factor of 1.9999(1) arising from 2D electrons at the Si/SiO$_2$ interface, similar to that reported previously~\cite{Shankar2008}. When V$_G$ is greater than the threshold voltage ($\mathrm{V}_{th} = \SI{1}{\volt}$) (the Fermi energy, E$_F$, above the conduction band edge, E$_C$), the signal corresponds to mobile 2D electrons in the conduction band. The signal magnitude is independent of V$_G$ above V$_{th}$, consistent with a constant density of states, g(E), of 2D electrons above E$_C$ (Fig.~\ref{fig:ESR signal vs gate voltage}(b)). On the other hand, when V$_G$ is below V$_{th}$ (i.e.\ E$_{F}$ $\le$ E$_{C}$), the ESR signal corresponds to electrons confined in a disorder induced band tail of states below E$_C$ (Fig.~\ref{fig:ESR signal vs gate voltage}(b) and Ref.~\onlinecite{Ando1982}, p. 517). The signal below V$_{th}$ exhibits a Curie susceptibility (inversely proportional to temperature) indicating that electrons in confined states behave as isolated spins\cite{Shankar2008}. Hence, gate voltage control of our MOSFET conveniently allows ESR measurements under identical conditions on both mobile 2D electrons as well as isolated electrons confined in quantum dots.

Figure~\ref{fig:ESR signal vs gate voltage}(a) illustrates a typical dependence of the ESR signal intensity, proportional to the number of unpaired electron spins (N$_{S}$) in the MOSFET, as a function of V$_{G}$. In these experiments, at each temperature, V$_{G}$ was first biased at \SI{2}{\volt} (i.e.\ above threshold) and then progressively reduced while measuring the spins. At \SI{4.8}{\kelvin} when measured in the dark (squares), N$_{S}$ decreases as V$_G$ decreases, and then becomes constant at $\mathrm{V}_{G} = \SI{0.4}{\volt}$ (marked as V$^*_{4.8K}$ in Fig.~\ref{fig:ESR signal vs gate voltage}(a)). For V$_{G}$ below \SI{0.4}{\volt} in the dark, electrons are confined into dots with a characteristic depth greater than k$_{B}$T (i.e.\ $\mathrm{E}_{C} - \mathrm{E}_{F} \gtrsim \mathrm{k}_{B} \mathrm{T}$, where k$_{B}$ is the Boltzmann constant and T is the temperature), and are therefore unable to thermally escape to the source or drain contacts. Brief illumination neutralizes the dots so that N$_{S}$ measured after illumination goes to zero at \SI{0}{\volt} (circles). In contrast, at a lower temperature of \SI{2}{\kelvin} (triangles), N$_{S}$ becomes constant at a higher V$_{G}$ of \SI{0.6}{\volt}; the dots have a correspondingly shallower depth ($\mathrm{E}_{C} - \mathrm{E}_{F} \gtrsim \SI{2}{\kelvin}$). Thus at any given temperature, by first biasing the gate above threshold and then reducing V$_{G}$ to \SI{0}{\volt}, we freeze electrons into natural quantum dots with a confinement depth characteristic of the temperature.

The characteristic gate voltages V$^*_{T}$ at which N$_{S}$ becomes constant were measured for several temperatures in the range from \SI{2}{\kelvin} to \SI{10}{\kelvin}. These V$^*_{T}$ were then used to estimate the energies and density of confined electron states below E$_C$. For each temperature, T, the total number of confined electrons was first calculated using the gate capacitance (\SI[per=slash]{2e11}{electrons\per\volt\cms}) and the corresponding measured V$^*_{T}$. Then, assuming that electrons are thermally activated, the characteristic confinement energy was calculated for that T\cite{[{The calculation is based on an approach developed in }][{. For details see }]Jenq1978,*Shankar2010}. Finally, the ratio of the difference in the numbers of confined electrons to the difference in the confinement energies between any two temperatures gives the density of confined electron states. Our result for the density of states is shown in Fig.~\ref{fig:ESR signal vs gate voltage}(c). Also shown for comparison is the density of states calculated for mobile 2D electrons at a (100) Si/SiO$_2$ surface (given by $g_v m / \pi \hbar^2$~\cite{Ando1982}, where $g_v$ = 2 is the valley degeneracy, $m$ is the effective mass of the electron, and $\hbar$ is the Planck's constant). As Fig.~\ref{fig:ESR signal vs gate voltage}(c) indicates, natural quantum dots in our sample have characteristic confinement energies of a few millivolts.

Having determined the confinement energies of the natural quantum dots, pulsed ESR spectroscopy as a function of gate voltage and temperature was performed to determine T$_1$ and T$_2$ for both natural dots and mobile electrons at the Si/SiO$_2$ interface. Most pulsed measurements on quantum dots were done with dots confined at \SI{5}{\kelvin}, i.e.\ V$_G$ was first taken above threshold at \SI{5}{\kelvin} and then brought back to \SI{0}{\volt} leaving only electrons trapped deeper than about \SI{4}{\meV} (Fig.~\ref{fig:ESR signal vs gate voltage}(c)). The device was then cooled to temperatures as low as \SI{350}{\milli\kelvin}, freezing electrons into these confined states. A single T$_2$ experiment was also done on dots confined at \SI{10}{\kelvin}, where the electrons are estimated to be trapped deeper than about \SI{11}{\meV}.

\begin{figure}
\centerline{\includegraphics*{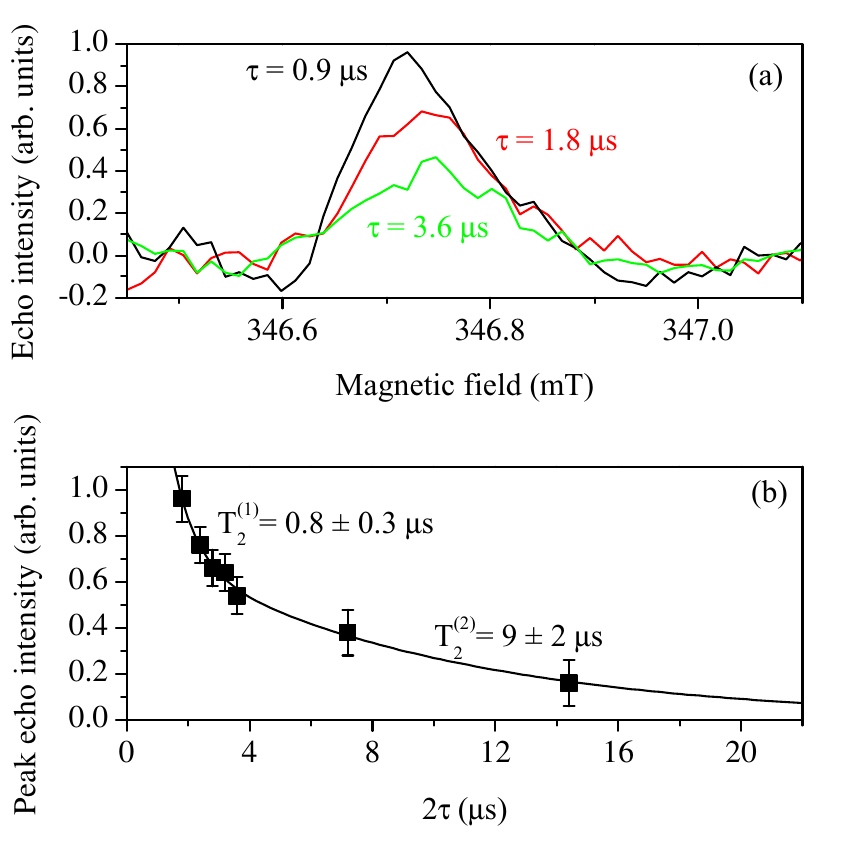}}
\caption{\label{fig:Spectra and T1,T2}T$_2$ experiment on electrons confined in dots at \SI{5}{\kelvin} as measured at \SI{350}{\milli\kelvin}. (a) Typical echo detected spectra after a two-pulse sequence for different $\tau$. (b) Peak intensity of the 2-pulse signal as function of $2\tau$. The bi-exponential fit gives two characteristic T$_{2}$'s.}
\end{figure} 

In the T$_2$ experiment, the magnetic field was swept during echo detection and a broad underlying baseline was subtracted in order to remove any echo signal contributed by background spins. Fig.~\ref{fig:Spectra and T1,T2}(a) shows typical field-sweep spectra, obtained after removing baselines, of the echo intensity at the end of a 2-pulse sequence. The signal at about \SI{366.7}{\milli\tesla} (g = 1.9999) belongs to electrons at the Si/SiO$_2$ interface. The peak intensity of the signal is plotted as a function of time in Fig.~\ref{fig:Spectra and T1,T2}(b), and the exponential fit gives T$_2$. In some experiments on natural dots, such as the one shown in Fig.~\ref{fig:Spectra and T1,T2}(b), the echo decays were non-exponential and best fitted using a bi-exponential function $\left(A_1\exp\left\{-2\tau/\mathrm{T}_2^{(1)}\right\}+A_2\exp\left\{-2\tau/\mathrm{T}_2^{(2)}\right\}\right)$.

\begin{figure}
	\centering
		\includegraphics{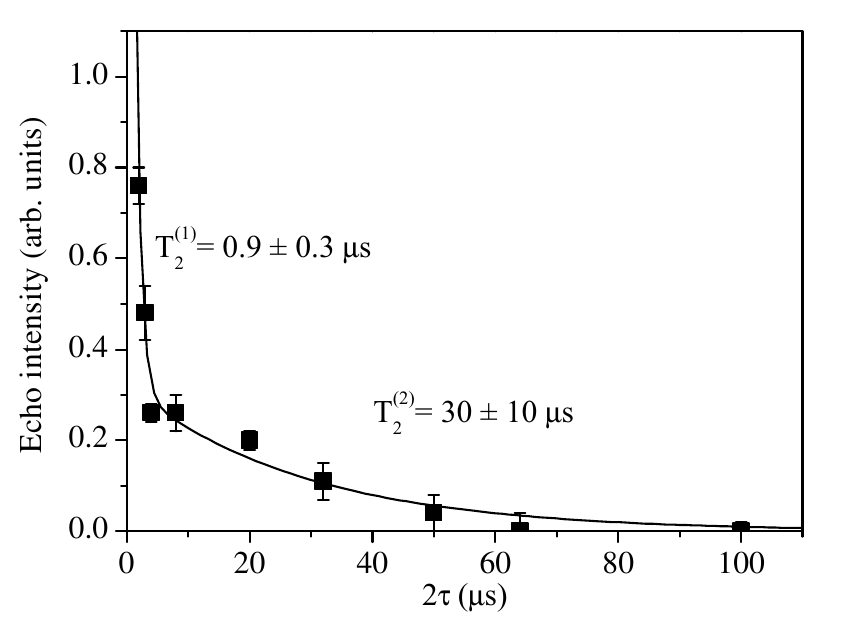}
	\caption{2-pulse echo decay for natural dots confined at \SI{10}{\kelvin}, measured at \SI{350}{\milli\kelvin}.}
	\label{fig:T2_10K_confined_dots}
\end{figure}

The bi-exponential decay arising from the natural dots is suggestive of a broad distribution of dots with a range of T$_2$ values. For example, for natural dots confined at \SI{5}{\kelvin} shown in Fig.~\ref{fig:Spectra and T1,T2}, approximately \SI{90}{\percent} of the electrons have a short T$_2$ of \SI{0.8}{\micro\second} while about \SI{10}{\percent} show a longer T$_2$ of \SI{9}{\micro\second}. A similar bi-exponential decay was observed for natural dots confined at \SI{10}{\kelvin} (Fig.~\ref{fig:T2_10K_confined_dots}); as discussed below, confinement at \SI{10}{\kelvin} reduces the total electron density by about a factor of two as compared to confinement at \SI{5}{\kelvin}. As seen in Fig.~\ref{fig:T2_10K_confined_dots}, the natural dots confined at \SI{10}{\kelvin} exhibit a similar fraction of dots showing a short T$_2$ of \SI{0.9}{\micro\second} while the remainder show an even longer T$_2$ of \SI{30}{\micro\second}. In Sec.~\ref{sec:Discussion}, we propose that the bi-exponential echo decay and the dependence of T$_2$ on the confinement temperature indicates that T$_2$ for these natural dots is controlled by exchange interactions between neighboring dots. Since the distance between any two dots is random, there will be a broad distribution in the exchange interaction between dots. The broad distribution in exchange results in some dots decohering quickly (a short T$_2$), while others decohere slowly (a long T$_2$); thus the echo signal cannot be characterized by a single exponential decay.

\begin{table}
\caption{Spin-counting results}
\label{tab:SpinCountingResults}
\begin{ruledtabular}
\begin{tabular}{ccc}
Confinement	& Density of dots		& Density of dots	\\
temperature	& in CW ESR			& that show long T$_2$	\\
(K)		& (\SI{e10}{\per\cms})		& (\SI{e9}{\per\cms})	\\
\hline
5		& 1.2				& 1.0			\\
10		& 0.7				& 0.5			\\
\end{tabular}
\end{ruledtabular}
\end{table}

We performed a spin-counting experiment in order to quantify the density of natural dots confined at \SI{5}{\kelvin} and \SI{10}{\kelvin}. These densities are used in Sec.~\ref{sec:Discussion} to estimate the magnitude of the exchange interactions between neighboring dots. In the spin-counting experiment, the absolute density of electron spins in dots measured in both CW and pulsed modes was obtained by comparing to the signal arising from the known thickness (\SI{25}{\micro\meter}) and density of isolated phosphorus donors (\SI{e14}{\per\cmc}) in the epi-layer substrate. We estimate that the error in the spin count is at most about a factor of two. The spin-counting results for signals from natural dots are tabulated in Table~\ref{tab:SpinCountingResults}. We find that the total density of confined dots, as well as dots having a long T$_2$, reduces by a factor of two upon changing the confinement temperature from \SI{5}{\kelvin} to \SI{10}{\kelvin}. Further, irrespective of the confining temperature, about \SI{10}{\percent} of all the dots seen in the CW experiment show a long T$_2$ in pulsed ESR at \SI{350}{\milli\kelvin}.

\begin{figure}
	\centering
		\includegraphics{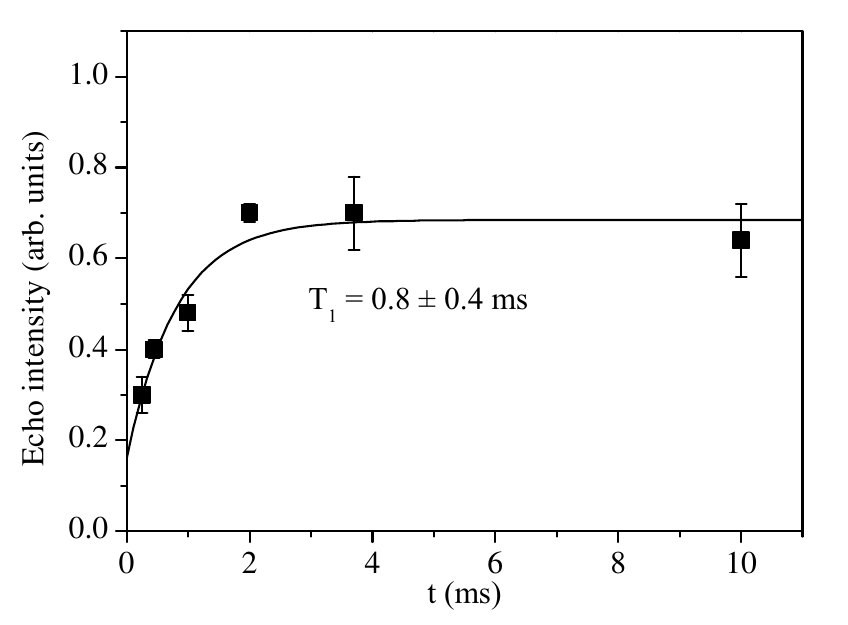}
	\caption{Echo intensity at the end of the 3-pulse experiment on natural dots confined at \SI{5}{\kelvin} and measured at \SI{350}{\milli\kelvin}. The exponential fit of the intensity as a function of t gives T$_{1}$.}
	\label{fig:T1}
\end{figure}

We measured the T$_1$ of only those dots that have a long T$_2$, by using a $\tau = \SI{1.6}{\micro\second}$ between the second and third pulses of the 3-pulse experiment, to ensure that all spins with $\mathrm{T}_2 = \SI{0.8}{\micro\second}$ have already decayed away. Currently, we have not measured the T$_1$ for electrons that showed a short T$_2$. To extract T$_1$, the echo intensity at the end of the 3-pulse sequence is plotted versus t and fit with an exponential recovery curve (Fig.~\ref{fig:T1}). Note that in an ideal inversion recovery experiment, the echo signal at short t (t $\le$ T$_1\ln(2)$) should be negative, corresponding to an inversion of the spins by the initial $\pi$ pulse. In our experiment, non-ideality of the inversion pulse due to the small microwave field B$_1$ ($\sim$ ESR linewidth) and its inhomogeneity across the sample, precludes us from observing a negative echo signal. Therefore, the intensity of the echo is fit by the exponential dependence $\left(a-b\exp\left\{-\mathrm{t}/\mathrm{T}_1\right\}\right)$ to find the characteristic time T$_1$.

\begin{figure}
\centerline{\includegraphics*{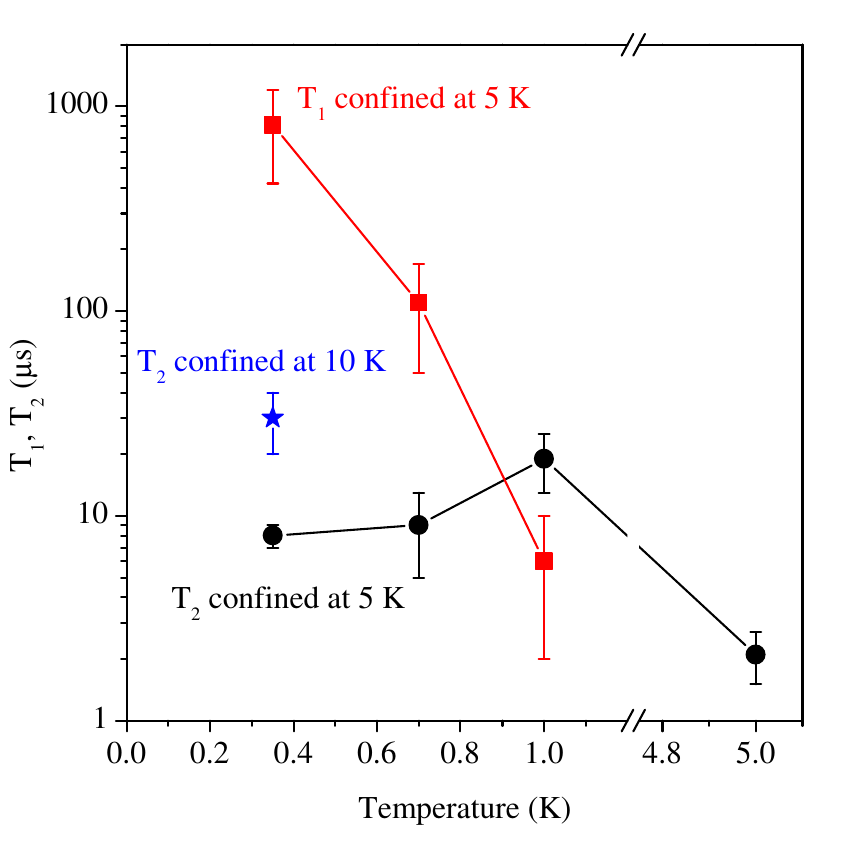}}
\caption{\label{fig:T1,T2 vs temperature}T$_{1}$ and T$_{2}$ as a function of temperature for natural dots that show a long T$_2$. Squares and circles correspond to natural dots confined at \SI{5}{\kelvin}, while the star shows the T$_2$ for dots confined at \SI{10}{\kelvin} and measured at \SI{350}{\milli\kelvin}. Lines are guides for the eye.}  
\end{figure} 

\begin{table}
\caption{\label{tab:T1, T2} Gate voltage dependence of T$_1$, T$_2$ and T$_2^*$.}
\begin{ruledtabular}
\begin{tabular}{ccccc}
V$_G$ (V)			& T (K)	& T$_1$ (\si{\micro\second})	& T$_2$ (\si{\micro\second}) 	& T$_2^*$ (\si{\micro\second})	\\
\hline
0 (confined at \SI{5}{\kelvin})	& 0.35	& 800 $\pm$ 400			& 8 $\pm$ 1			& 0.3				\\
2 (mobile)			& 5.0	& 0.33 $\pm$ 0.05		& 0.39 $\pm$ 0.04		& 0.3				\\
\end{tabular}
\end{ruledtabular}
\end{table} 

Table~\ref{tab:T1, T2} summarizes the gate voltage dependence of T$_1$, T$_2$ and T$_2^*$ contrasting natural quantum dots confined at \SI{5}{\kelvin} with mobile electrons. Fig.~\ref{fig:T1,T2 vs temperature} summarizes the temperature dependence of T$_1$'s and T$_2$'s of natural dots when confined at \SI{5}{\kelvin} and \SI{10}{\kelvin}.

\section{Discussion}
\label{sec:Discussion}

Mobile electrons ($\mathrm{V}_G = \SI{2}{\volt}$) have sub-microsecond T$_1$ and T$_2$ at \SI{5}{\kelvin}, an order of magnitude shorter than the relaxation times measured for 2D electrons in high-mobility Si/SiGe heterostructures~\cite{Tyryshkin2005}. As in Si/SiGe structures, T$_1$ and T$_2$ relaxation may be caused by a fluctuating Rashba effective magnetic field arising from the spin-orbit coupling~\cite{BychkovRashba1984,Tahan2005}. Alternatively, other spin relaxation processes also arising from spin-orbit coupling such as the Elliot-Yafet mechanism~\cite{Elliott1954,Yafet1963,Cheng2010} could result in short T$_1$'s and T$_2$'s especially in low mobility structures~\cite{Wilamowski2004} like a MOSFET. While a detailed understanding of the process causing spin relaxation of mobile 2D electrons remains a topic for future work, for now spin relaxation can be plausibly associated with spin-orbit coupling modulated by mobile electron scattering.

Upon confining 2D electrons into quantum dots and cooling to low enough temperature, theory~\cite{Khaetskii2001} predicts that the Rashba field should become less effective in inducing relaxation and therefore longer T$_1$ and T$_2$ should emerge. In agreement, our results as summarized in Table~\ref{tab:T1, T2} ($\mathrm{V}_G = \SI{0}{\volt}$) and Fig.~\ref{fig:T1,T2 vs temperature} show that T$_1$ for natural dots confined at \SI{5}{\kelvin} rises with decreasing temperature reaching \SI{0.8}{\milli\second} at \SI{350}{\milli\kelvin}. While the detailed functional form of T$_1$ seen in Fig.~\ref{fig:T1,T2 vs temperature} is unclear, the fact that T$_1$ rises as temperature decreases is in line with trends expected for phonon related mechanisms. Our results are qualitatively in agreement with recent experiments on small (\SI{200}{\nm} lithographic dimensions) dots in silicon structures~\cite{Hayes2009,Xiao2010}, demonstrating that T$_1$ is controlled by inelastic phonon scattering and that this scattering can be suppressed at lower temperatures, leading to T$_1 \sim \SI{1}{\second}$ at \SI{48}{\milli\kelvin}. More temperature points and a study of T$_1$ as a function of confinement energy need to be added to the data in Fig.~\ref{fig:T1,T2 vs temperature} in order to make quantitative comparisons to the theory and other experiments. However for now, the T$_1$ data show that by confining 2D electrons into quantum dots with a few millivolts binding energy and cooling to low enough temperatures (\SI{350}{\milli\kelvin}), Rashba field fluctuations are made less effective and T$_1$ can be increased by almost four orders of magnitude.

Restricting electron motion also increases T$_2$ (Fig.~\ref{fig:T1,T2 vs temperature}). After confining into quantum dots at \SI{5}{\kelvin}, the spins have a T$_2$ of around \SI{2}{\micro\second} at \SI{5}{\kelvin}, which is about five times longer than mobile 2D electrons. Upon cooling, T$_2$ increases to around \SI{20}{\micro\second} at \SI{1}{\kelvin} where as seen in Fig.~\ref{fig:T1,T2 vs temperature}, it is comparable to T$_1$. Therefore at \SI{1}{\kelvin}, T$_2$ is limited by the energy relaxation mechanisms that control T$_1$ ($\mathrm{T}_2 \sim \mathrm{T}_1$). T$_2$ appears to decrease on further cooling to around \SI{10}{\micro\second} at \SI{350}{\milli\kelvin} (error bars are \SI{67}{\percent} confidence bounds). Significantly, T$_2$ is two orders of magnitude less than T$_1$ at \SI{350}{\milli\kelvin}. The fact that $\mathrm{T}_2 \ll \mathrm{T}_1$ at \SI{350}{\milli\kelvin} demonstrates that T$_2$ is not limited by T$_1$ relaxation but instead the spins decohere by a different, extrinsic mechanism.

We can rule out fluctuating hyperfine interactions due to nuclear spins~\cite{Hanson2007} as the extrinsic mechanism limiting coherence at \SI{350}{\milli\kelvin}. This extrinsic process limits the Hahn echo T$_2$ in individual GaAs quantum dots from a few \si{\micro\second}~\cite{Petta2005,Koppens2008} to \SI{30}{\micro\second}~\cite{Bluhm2010}, much shorter than T$_1$. However, similar processes cannot explain the \SI{10}{\micro\second} T$_2$ measured in our device since it is fabricated on an isotopically enriched $^{28}$Si substrate having a negligible number of nuclear spins (\SI{800}{ppm} $^{29}$Si fraction).

A clue to the mechanism controlling T$_2$ is provided by the apparent increase in T$_2$ from \SI{350}{\milli\kelvin} to \SI{1}{\kelvin}. Such an increase of T$_2$ with temperature is reminiscent of motional narrowing of a decoherence mechanism involving interactions between spins. The exchange coupling provides one possible mode for such an interaction. Simple estimates of the wavefunction overlap and the exchange interaction between neighboring dots can be made for electrons confined at \SI{5}{\kelvin} where the confinement energy is at least about \SI{4}{\meV} (Fig.~\ref{fig:ESR signal vs gate voltage}(c)) and the distance between dots is about \SI{90}{\nm} (dot density of \SI{1.2e10}{\per\cms}). With these parameters and a simple quartic confining potential\cite{Burkard1999}, one calculates an exchange of about \SI{300}{\kHz}, in the same range as $1/\mathrm{T}_2$. Precise agreement is not expected as the exchange is exponentially sensitive to the dot spacing and depth. This idea that exchange between neighboring dots controls T$_2$ is also supported by the longer T$_2$ observed when the dots are confined at \SI{10}{\kelvin} (Fig.~\ref{fig:T1,T2 vs temperature}) since the lower electron density implies a larger distance between dots and consequently smaller exchange coupling. Further, in practice there will be a broad distribution of exchange between dots in the sample; this distribution in exchange could account for the variation in T$_2$ seen in Fig.~\ref{fig:Spectra and T1,T2}(b) and Fig.~\ref{fig:T2_10K_confined_dots}. A fluctuating exchange between dots appears to be the likely extrinsic decoherence mechanism that is causing $\mathrm{T}_2 \ll \mathrm{T}_1$.

\section{Conclusion}
\label{sec:conclusion}

We have shown that for electrons confined in natural quantum dots having a few millivolts of confinement energy, T$_1$ reaches almost a millisecond and the longest T$_2$ is about \SI{30}{\micro\second} at \SI{350}{\milli\kelvin}. On the other hand, mobile 2D electrons have short T$_1$ and T$_2$ of about \SI{0.3}{\micro\second}, possibly controlled by a Rashba field. Our results confirm that confinement makes Rashba fluctuations less effective leading to a four order of magnitude longer T$_1$ in isolated quantum dots at the Si/SiO$_2$ interface, similar to GaAs quantum dots and many-electron dots in Si/SiGe and Si/SiO$_2$ heterostructures. Recently there has been encouraging progress towards a scalable quantum computing architecture using gated dots in Si/SiGe~\cite{Simmons2009} and Si/SiO$_2$~\cite{Nordberg2009} systems. The long T$_1$ we measure in silicon quantum dots suggest that these schemes have a promising future.

While the natural dots show a T$_2$ about one to two orders of magnitude longer than mobile electrons, the T$_2$ is still significantly shorter than T$_1$. This shorter T$_2$ is likely due to exchange interactions between neighboring dots, an explanation which is consistent with estimates and the experimental observation of a threefold increase in T$_2$ at lower electron density. The lowest electron density (\SI{0.5e9}{\per\cms}) in this work is at the limit of the sensitivity of our ESR spectrometer. Experiments which eliminate the exchange effects will require improved sensitivity.

\begin{acknowledgments}
This work was supported by the NSF through the Princeton MRSEC (DMR-0819860) and by NSA/LPS and ARO through the University of Wisconsin (W911NF-08-1-0482).
\end{acknowledgments}

\bibliography{pulsed_ESR_confined_electrons_paper}

\end{document}